\documentstyle[amssymb,amsmath,epsfig]{article}

\begin{document}
%\draft

\title{Constraints on the generalized Chaplygin gas from supernovae observations}
\author{Mart\'\i n Makler$^{1,2}$, S\'{e}rgio Quinet de Oliveira$^{2}$ and Ioav
Waga$^{2}$\\
 $^{1}$Centro Brasileiro de Pesquisas F\'\i sicas\\
Rua Xavier Sigaud, 150, CEP 22290-180 Rio de Janeiro, RJ,
Brazil\\
$^{2}$Universidade Federal do Rio de Janeiro, \\
Instituto de F\'\i sica, \\
CEP 21945-970 Rio de Janeiro, RJ, Brazil}
\date{\today}
\maketitle

\begin{abstract}
We explore the implications of type Ia supernovae (SNIa)
observations on flat cosmological models whose matter content is
an exotic fluid with equation of state, $p=-M^{4(\alpha
+1)}/\rho^{\alpha}$. In this scenario, a single fluid component
may drive the Universe from a nonrelativistic matter dominated
phase to an accelerated expansion phase behaving, first, like dark
matter and in a more recent epoch as dark energy. We show that
these models are consistent with current SNIa data for a rather
broad range of parameters. However, future SNIa experiments will
place stringent constraints on these models, and could safely rule
out the special case of a Chaplygin gas ($\alpha=1$) if the
Universe is dominated by a true cosmological constant.
\end{abstract}

\section{Introduction}

According to the standard cosmological scenario ($\Lambda $CDM,
QCDM) that has emerged at the end of the last century, the
Universe is dominated by two unknown components with quite
different properties: pressureless cold dark matter (CDM), which
is responsible for the formation of structures, and
negative-pressure dark energy, that powers the accelerated
expansion. There are several candidates for these two components.
For the CDM, the leading particle candidates are the axion and the
neutralino, two weakly interacting massive particles. The
preferred candidates for dark energy are vacuum energy - or a
cosmological constant $\Lambda $ - and a dynamical scalar field
(quintessence) \cite{denergy}. At the cosmological level, the
direct detection of each of these two components involves
observations at different scales. Since it is not supposed to
cluster at small scales, the effect of dark energy can only be
detected over large distances, where the accelerated expansion is
observed. On the other hand, the CDM can be detected by its local
manifestation on the motion of visible matter or through the
bending of light in gravitational lensing.

An interesting question that arises is: could this two phenomena -
accelerated expansion and clustering - be different manifestations of a
single component? In principle the answer is yes, if, for instance, the
Universe is dominated by a component with an appropriate exotic equation of
state (EOS). We will generically refer to any kind of such unifying dark
matter-energy component as {UDM}. \footnote{%
Following the current jargon, another possible denomination for UDM would be
``quartessence'' since in this scenario we have only one additional
component, besides ordinary matter, photons and neutrinos, and not two like
in $\Lambda $CDM and QCDM.}

The above question has been addressed in some works recently \cite
{wetterich,bilic,davidson,padmanabhan}. For instance, in Ref. \cite
{padmanabhan} it was investigated the possibility that a tachyonic field,
with motivation in string theory, could unify dark energy and dark matter
and explain cosmological observations in small and large scales. Here we
investigate observational limits on a simple realization of UDM: a fluid
with the following equation of state \cite{bilic,makler,bento,kamenshchik},
\begin{equation}
p=-\frac{M^{4(\alpha +1)}}{\rho ^{\alpha }}.  \label{eostoy}
\end{equation}
The particular case $\alpha =1$ is known as Chaplygin gas and its
cosmological relevance, as an alternative to quintessence, has been pointed
out in \cite{kamenshchik}. In \cite{bilic}, it has been shown that the
inhomogeneous Chaplygin gas represents a promising model for dark
matter-energy unification. Some possible motivations for this scenario from
the field theory point of view are discussed in \cite
{kamenshchik,bilic,bento}. The Chaplygin gas appears as an effective fluid
associated with $d$-branes \cite{kamenshchik,bazeia}. The same EOS is also
derived from a complex scalar field with appropriate potential and from a
Born-Infeld Lagrangian \cite{bilic}. More recently, by extending the work of
Bili\'{c} {\it et al.} \cite{bilic}, Bento {\it et al.} \cite{bento} also
discussed the particle physics motivation for the EOS (\ref{eostoy}). The
fluid given by this EOS is sometimes called generalized Chaplygin gas (GCG).

It is interesting to notice that this model can also be obtained from purely
phenomenological arguments, by requiring that an exotic fluid unifies the
dark-matter/dark-energy behavior as a function of its density and that it is
stable and causal \cite{makler}. The simplest EOS satisfying this criteria
is given by eq. (\ref{eostoy}).

Let us consider the homogeneous case of the GCG Universe. The energy
conservation can be written as
\begin{equation}
d\rho=-3\left( \rho + p\right)\frac{da}{a},
\end{equation}
where $a$ is the scale factor. By solving this equation we may express the
energy density in terms of the scale factor:
\begin{equation}
\rho = M^4 \left[ B \left(\frac{a_0}{a}\right)^{3(\alpha +1)}+1 \right]
^{1/( \alpha +1) },  \label{rhotoy}
\end{equation}
where $a_0$ is present value of scale factor and $B$ is an integration
constant. When $a/a_0\ll 1$, we have $\rho \propto a^{-3}$ and the fluid
behaves as CDM. For late times, $a/a_0\gg 1$, and we get $p=-\rho =-M^4
=const.$ as in the cosmological constant case. There is also an intermediate
phase where the effective EOS is $p=\alpha \rho$ \cite{kamenshchik}. Once we
have $\rho$ as a function of the scale factor it is simple to find the
Hubble parameter. Since observations of anisotropies in the cosmic microwave
background (CMB) indicate that the Universe is nearly flat \cite{balbi},
here we restrict our attention to the zero curvature case. We also neglect
radiation, that it is not relevant for the cosmological tests we discuss in
this work.

>From the Friedmann equation with $k=0$ we have
\begin{equation}
H^{2}\left( z\right) =H_{0}^{2}\left[ \Omega _{M}^*\left(
1+z\right) ^{3\left( \alpha +1\right) }+\left( 1-\Omega
_{M}^*\right) \right] ^{1/\left( \alpha +1\right) },
\end{equation}
where $z=a_{0}/a-1$ is the redshift, and we have conveniently
defined $ \Omega _{M}^*=B/(B+1)$, or equivalently
\begin{equation}
B=\frac{\Omega _{M}^*}{1-\Omega _{M}^*}.  \label{B}
\end{equation}
Further, we also have
\begin{equation}
M^{4}=\rho _{c0}\left( 1-\Omega _{M}^*\right) ^{\frac{1}{\left(
\alpha +1\right) }},  \label{m4}
\end{equation}
where $\rho _{c0}$ is the present value of the critical density.
For these models the deceleration parameter can be written as
\begin{equation}
q=-\frac{\stackrel{.}{H}}{H^{2}}-1=\frac{\frac{\Omega
_{M}^*}{2}-\left( 1-\Omega _{M}^*\right) \left( 1+z\right)
^{-3(1+\alpha )}}{\Omega _{M}^*+\left( 1-\Omega _{M}^*\right)
\left( 1+z\right) ^{-3(1+\alpha )}},
\end{equation}
and the redshift $z_{\ast }$, at which the Universe started its accelerating
phase is given by,
\begin{equation}
1+z_{\ast }=\left( \frac{2\;(1-\Omega _{M}^*)}{\Omega
_{M}^*}\right) ^{\frac{1 }{3(\alpha +1)}}.
\end{equation}
An accelerating Universe at present time ($q_{0}<0$) implies that
$\Omega _{M}^*<2/3$, and from (\ref{B}) we have $0<B<2$; the lower
limit follows from the fact that we assume $\Omega _{M}^*>0$.
Moreover, if $\alpha $ is not very close to $-1$, from (\ref{m4}),
we obtain $M\sim 10^{-3}$ eV. It would be desirable that a
fundamental theory, aimed to describe the UDM, sheds some light on
the origin of this mass scale. Thus, at this point this model is
not free of some tuning. However, once the origin of the above
mass scale is explained, the so called dark matter-energy
``coincidence problem'' is not present in this scenario.

In a GCG Universe, if the parameter $\alpha $ is positive, the adiabatic
sound velocity, ${c_{s}}^{2}=dp/d\rho =-\alpha p/\rho $, is real and
therefore, the fluid component is stable. If $\alpha $ is negative and there
is only adiabatic pressure fluctuations, they accelerate the collapse
producing instabilities that turn the model for structure formation
unacceptable \cite{fabris1,hu}. Moreover, to obey causality, the sound
velocity in this medium has to be less or equal than the light velocity.
Since the maximum allowed sound velocity of this fluid (which occurs in the
regions where $p\rightarrow -\rho $) is given by $\sqrt{\alpha }$, this
condition imposes $\alpha \leq 1$. The Chaplygin gas, $\alpha =1$, is the
extreme case, where the sound velocity can be nearly the speed of light. The
case $\alpha =0$ is equivalent to $\Lambda $CDM and is, of course, well
motivated. In this paper, we discuss the GCG model from a phenomenological
point of view. Hence, although we are aware that most likely $0\leq \alpha
\leq 1$, we also include in our analysis the region where $\alpha $ is
negative, but larger than $-1$. If $\alpha =-1$ we obtain a de Sitter
Universe. The situation $\alpha <-1$ seems unphysical, since the energy
density of UDM would be increasing with the expansion of the Universe. In
fact, as we shall see, age constraints can safely exclude regions in the
parameter space with very negative values of $\alpha $.

In the forthcoming section we will see what constraints to the model
described above are set by present and future SNIa observations. Recently,
some constraints from SNIa on related models where obtained in Ref. \cite
{fabris2}. The work presented here differs from \cite{fabris2} in the
following aspects: a) Following the idea of unification, we have not
included an additional dark matter component and we have considered the more
general case in which $\alpha $ is not necessarily equal to unity. b) When
analyzing current SNIa data we perform a Bayesian approach in which the
intercept is marginalized c) We also investigate the predicted constraints
on the models from future SNIa observations.

\section{Type Ia Supernovae Experiments}

The luminosity distance of a light source is defined in such a way as to
generalize to an expanding and curved space the inverse-square law of
brightness valid in a static Euclidean space,

\begin{equation}
d_{L}=\left( \frac{L}{4\pi {\cal F}}\right)
^{1/2}=\,\,(1+z)\,\int_{0}^{z}\, \frac{dy}{H(y)}.  \label{L}
\end{equation}
In (\ref{L}) ${L}$ is the absolute luminosity and ${\cal F}$ is the measured
flux.

For a source of absolute magnitude $M$, the apparent bolometric
magnitude $ m(z)$ can be expressed as
\begin{equation}
m(z)={\cal M}+5\;\log D_{L},  \label{appmag}
\end{equation}
where $D_{L}=D_{L}(z,\alpha ,\Omega _{M}^*)$ is the luminosity
distance in units of $H_{0}^{-1}$, and
\begin{equation}
{\cal M}=M-5\;\log H_{0}+25
\end{equation}
is the ``zero point'' magnitude (or Hubble intercept magnitude).

In our computations we follow the Bayesian approach of Drell, Loredo and
Wasserman \cite{drell} (see also \cite{ng}) and we direct the reader to
these references for details. We consider the data of fit C, of Perlmutter
{\it \ et al.} \cite{perlmutter}, with 16 low-redshift and 38 high-redshift
supernovae. In our analysis we use the following marginal likelihood,

\begin{equation}
{\cal L} (\alpha ,\Omega _{M}^*)=\frac{s\sqrt{2\pi }}{\Delta \eta
}\,e^{- \frac{q}{2}}.
\end{equation}
Here

\begin{eqnarray}
q(\alpha ,\Omega _{M}^*)
&=&\sum\limits_{i=1}^{16}\frac{(-5\,\text{log}
D_{L}-n_{i}+m_{Bi}^{corr})^{2}}{\sigma _{low,i}^{2}}+  \nonumber \\
&&\sum\limits_{i=1}^{38}\frac{\left( -5\,\text{log}D_{L}-n_{i}+m_{Bi}^{eff}
\right) ^{2}}{\sigma _{high,i}^{2}},
\end{eqnarray}

where

\begin{eqnarray}
n_{i}(\alpha ,\Omega _{M}^*)
&=&s^{2}(\,\sum\limits_{i=1}^{16}\frac{5\,\text{
log}D_{L}(z_{i},\alpha ,\Omega _{M}^*)-m_{Bi}^{corr}}{\sigma
_{low,i}^{2}}+
\nonumber \\
&&\sum\limits_{i=1}^{38}\frac{5\,\text{log}D_{L}(z_{i},\alpha
,\Omega _{M}^*)-m_{Bi}^{eff}}{\sigma _{high,i}^{2}}\,)\,,
\end{eqnarray}

\begin{equation}
s^{2}=\left( \sum\limits_{i=1}^{16}\frac{1}{\sigma _{low,i}^{2}}
+\sum\limits_{i=1}^{38}\frac{1}{\sigma _{high,i}^{2}}\right) ^{-1},
\end{equation}

\begin{equation}
\sigma _{low,i}^{2}=\sigma _{m_{B,i}^{corr}}^{2}+\left( \frac{5\,\text{
log\thinspace }e}{z_{i}}\;\sigma _{z_{i}}\right) ^{2}
\end{equation}
and

\begin{equation}
\sigma _{high,i}^{2}=\sigma _{m_{B,i}^{eff}}^{2}+\left( \frac{5\,\text{
log\thinspace }e}{z_{i}}\;\sigma _{z_{i}}\right) ^{2}.
\end{equation}
The quantities $m_{B}^{corr}$, $m_{B}^{eff}$, $\sigma
_{m_{B}^{corr}}$, $ \sigma _{m_{B}^{eff}}$ and $\sigma _{z}$ are
given in Tables 1 and 2 of Perlmutter{\it \ et al.}
\cite{perlmutter}.

The results of our analysis for the GCG Universe are displayed in
Fig. $1$. In this figure we show $68$ and $95$ confidence level
contours, in the $ (\alpha ,$ $\Omega _{M}^*)$-plane. We observe
that current SNIa data constrain $\Omega _{M}^*$ to the range
$0.15\lesssim \Omega _{M}^*\lesssim 0.4$ , but do not strongly
constrain the parameter $\alpha $ in the considered range. Other
tests may impose further constraints. For instance, in Ref.
\cite{knox} it shown that CMB alone, imposes $T_{0}=14\pm 0.5$ Gyr
($1\sigma $) for the age of the Universe. If we also assume the
HST Key Project result, $H_{0}=72\pm 8$ \cite{freedman}, and that
$H_{0}$ and $T_{0}$ measurements are uncorrelated, we obtain for
the product $H_{0}T_{0}$, the following range:
$0.79<H_{0}\;T_{0}<1.27$, at the $2\sigma $ confidence level. The
central value occurs at $H_{0}T_{0}=1.03$. In Fig. 1, we also
display the contours $H_{0}T_{0}=0.79$ and $H_{0}T_{0}=1.27$. As
remarked before, we can see that negative values of $\alpha $
close to $-1$, are disfavored. We have also checked that, keeping
all other parameters fixed,  the position of the first Doppler
peak decreases as $\alpha$ increases. It would be interesting to
investigate the constraints imposed by cosmic microwave
observations on GCG models, but we leave this for future work.

Finally we consider how well the proposed Supernova Acceleration
Probe (SNAP) \cite{snap}, may constrain the parameters $\alpha $
and $\Omega _{M}^*$ . Following previous investigations
\cite{sne}, we assume, in our Monte Carlo simulations, that a
total of 2000 supernovae (roughly one year of SNAP observations)
will be observed with the following redshift distribution. We
consider, 1920 SNIa, distributed in 24 bins, from $z=0$ to
$z=1.2$. From redshift $z=1.2$ to $z=1.5$, we assume that 60 SNIa
will be observed and we divide them in 6 bins. From $z=1.5$ to
$z=1.7$ we consider 4 bins with 5 SNIa in each bin. All the
supernovae are assumed to be uniformly distributed with $\Delta
z=0.05$. In our simulations, we assume that the errors in
magnitude are Gaussian distributed with zero mean and variance
$\sigma _{m}=0.16$. This includes observational errors and
intrinsic scatter in the SNIa absolute magnitudes. We neglect, in
our simulations, uncertainties in the redshift. We also
investigated the effect of a redshift dependent systematic error
of the kind $\delta m=\pm (0.02/1.5)\;z$. This kind of systematic
error slightly shifts the ``ellipses'' up or down - depending if
the signal in $\delta m$ is plus or minus - but not along the
major axis of the ``ellipses''. We have not considered in this
work the systematic effect of lensing \cite{lens1}. This important
effect, is not expected to change qualitatively our conclusions,
unless the Universe contains a significant fraction of compact
objects \cite{lens2}. In this case, a more detailed analysis is
required \cite{quinet}.

\begin{figure} \centering \hspace*{0.in}
\epsfig{file=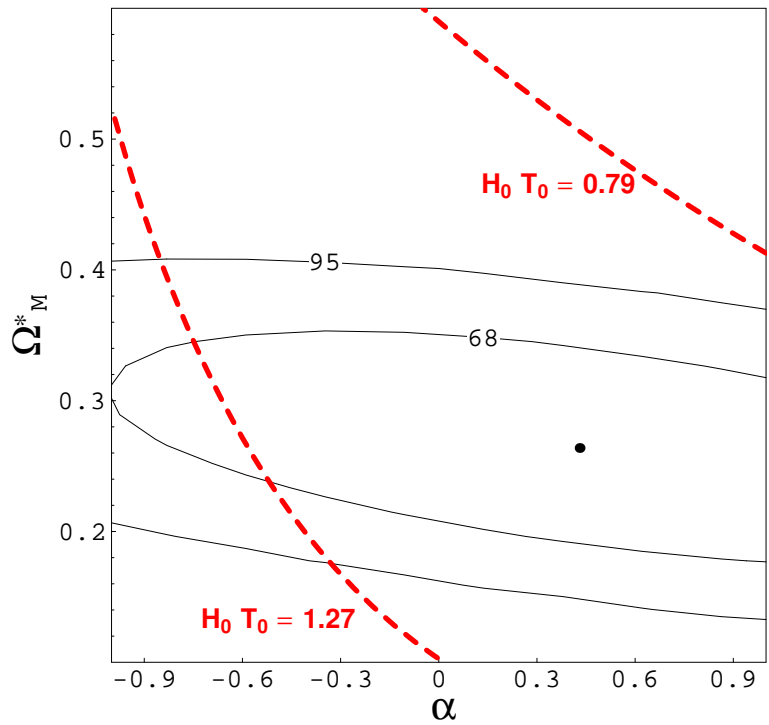,height= 10.0 cm,width= 10.0 cm}
 \vspace*{0.1in}\caption{In the figure $68$ and
$95$ confidence level contours, in the $(\alpha $, $\Omega
_{M}^*)$-plane, are displayed. For the figure we use fit C, of
Perlmutter \textit{et al.} \protect\cite{perlmutter}. The point in
the figure, with coordinates ($0.43, 0.26$), represents the best
fit value. Constraints from the age of the Universe give $0.79 <
H_0\;T_0 < 1.27$ (at the $2\sigma$ confidence level), the dashed
lines represent these two limits.}
\end{figure}

In Fig. 2 we display the results of our simulation assuming a
fiducial model with $\Omega _{M}^*=0.3$ and $\alpha =0$. For the
figure the Hubble intercept is assumed to be exactly know. In Fig.
3, we considered the case in which the intercept ${\cal M}$ is not
known, and we marginalized over it following Goliath {\it et. al.}
\cite{sne}. In Fig. 4 the fiducial model has $ \Omega _{M}^*=0.3$
and $\alpha =1$, and again the intercept is not assumed to be
known. From the figures it is clear that SNAP will be able to rule
out the Chaplygin gas model ($\alpha=1$) if the Universe is
dominated by a true cosmological constant. Alternatively, if the
Universe is dominated by the Chaplygin gas a cosmological constant
can be ruled out.

\begin{figure*}
\hspace*{0.1in} \psfig{file=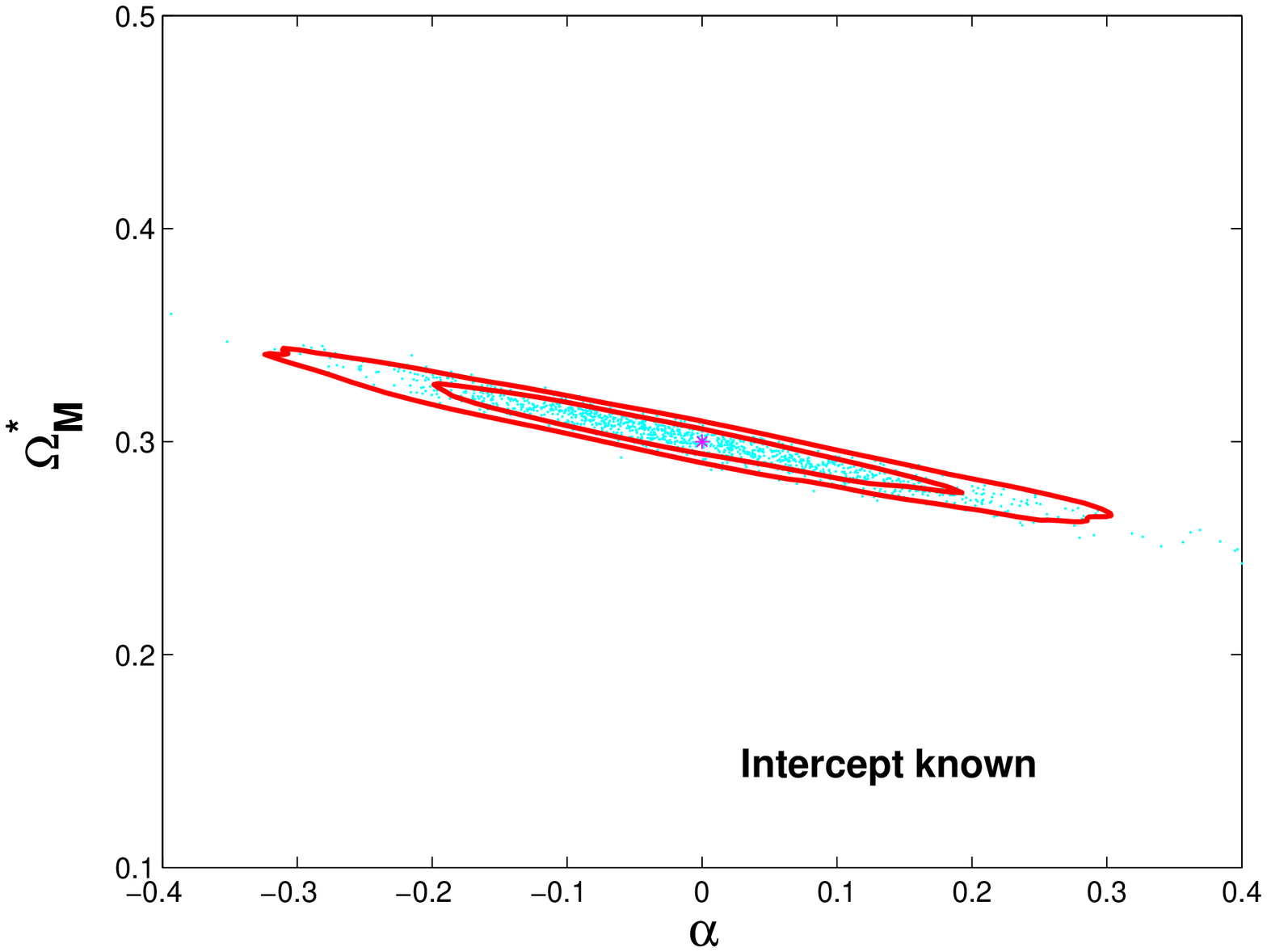,height=10.0cm,width=10.0cm}
\vspace*{0.1in} \caption{Predicted 68 and 95 confidence level
contours for the SNAP mission are shown. We considered a fiducial
model with $\Omega _{M}^*=0.3$ and $\alpha =0$. For the figure the
Hubble intercept is supposed to be known.}
\end{figure*}

\begin{figure*}
\hspace*{0.1in} \psfig{file=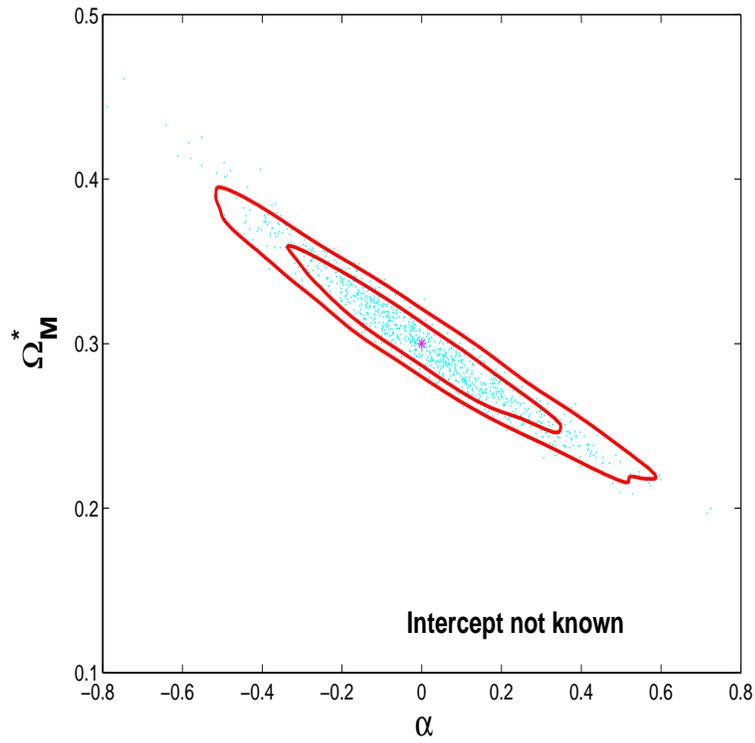,height=10.0cm,width=10.0cm}
\vspace*{0.1in} \caption{Predicted 68 and 95 confidence level
contours for the SNAP mission are shown. We considered a fiducial
model with $\Omega _{M}^*=0.3$ and $\alpha =0$. For the figure the
Hubble intercept is not supposed to be known.}
\end{figure*}

\section{Summary}

We derived constraints, from current and future SNIa observations,
in a scenario where both the accelerated expansion and CDM are
manifestations of a single component. We considered the special
case of a generalized Chaplygin gas. For the homogenous model, an
important difference between UDM and models with $\Lambda $ or
scalar fields is that in the former there is a transformation of
effective CDM into effective dark energy that produces the
accelerated expansion.

Our results show that the GCG is consistent with current SNIa data, for any
value of $\alpha $ in the considered range, although values of $\alpha \sim
0.4$ are favored. If the accelerated expansion is caused by a cosmological
constant, than SNAP data should be able to rule out the Chaplygin ($\alpha
=1 $) gas and alternatively, if the Universe is dominated by the Chaplygin
gas a cosmological constant would be ruled out with high confidence.

For simplicity, we have discussed in this letter the case of a
Universe composed of UDM only. Of course, one should also include
the baryonic component, whose energy density scales differently
from the UDM. When baryons are included in the Hubble parameter
the picture does not change, although some details do. For
instance, if we introduce $\Omega_b$ and perform the analysis with
the current supernovae data, the results for $\Omega _{M}^*$ stay
almost unchanged, but the best fit value for $\alpha$ decreases
($\alpha \sim 0.15$ for $\Omega_b \sim 0.04$, instead of $\alpha
\sim 0.4$ for $\Omega_b = 0$). Also, the age constraints on
$\alpha$ are weaker. For instance, for $\Omega_b = 0.04$ we can
exclude negative values of $\alpha$ close to $-1$ only for $\Omega
_{M}^*\lesssim 0.3$. In the case of the data expected from SNAP,
we redid the analysis of the preceding section for $\Omega_b=0.04
\pm 0.004$, assuming a Gaussian distribution. We marginalized over
$\Omega_b$ and noticed that the contours increase only slightly.

The GCG seems to be a promising model for unifying dark matter and
dark energy. More generically, the idea of UDM (``quartessence'')
has to be explored further, both from the particle physics point
of view - to provide a fundamental theory to it -, as well as from
the observational side, to constrain UDM models guiding us to
unveil its nature.

\bigskip {\bf Note added:} After this manuscript was submitted for 
publication, another paper using the GCG and SNIa obervations appeared 
on the web \cite{avelino}. Their results are similar to ours, although 
they do not set constraints on the parameter $\alpha$ of the GCG 
equation of state.

\begin{figure*}
\hspace*{0.1in} \psfig{file=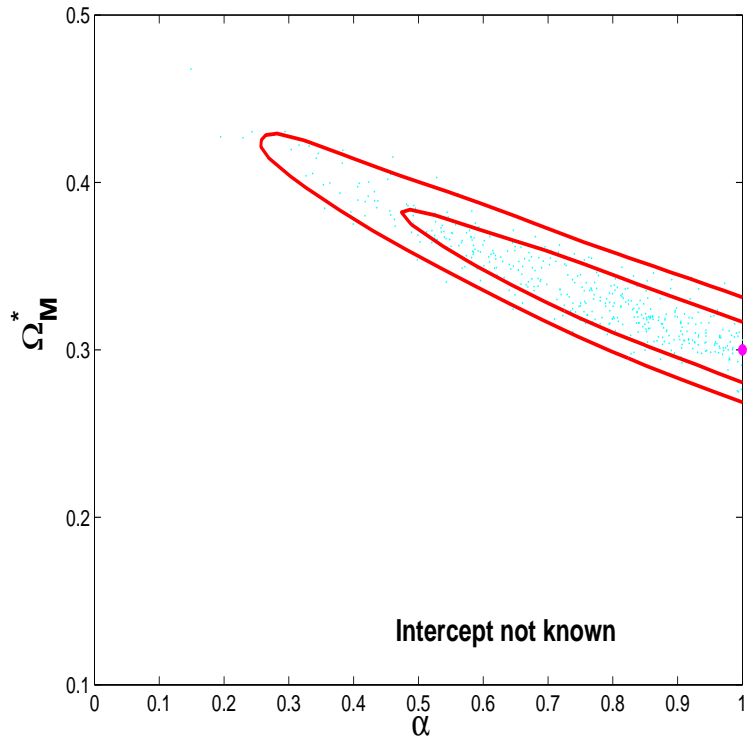,height=10.0cm,width=10.0cm}
\vspace*{0.1in} \caption{Predicted 68 and 95 confidence level
contours for the SNAP mission are shown. We considered a fiducial
model with $\Omega _{M}^*=0.3$ and $\alpha =1$. For the figure the
Hubble intercept is not supposed to be known.}
\end{figure*}

\bigskip {\bf Acknowledgments}

IW is partially supported by the Brazilian research agencies CNPq and
FAPERJ. SQ is partially supported by CNPq. MM wishes to acknowledge the
hospitality of Fermilab.

\end{document}